\documentclass[twocolumn,showpacs,preprintnumbers,amsmath,amssymb,prx]{revtex4-1}

\usepackage[english]{babel}
\usepackage{color}
\usepackage[usenames,dvipsnames,svgnames,table]{xcolor}
\usepackage{verbatim}

\usepackage{amsmath}
\usepackage{amssymb}

\newcommand{\ket}[1]{|#1\rangle}
\newlength{\eqboxstorage}

\usepackage{graphicx}
\usepackage{dcolumn}
\usepackage{bm}

\begin{document}

\title{Theoretical models of Rashba spin splitting in asymmetric  SrTiO$_3$-based heterostructures}

\author{L. W. van Heeringen$^1$, A. McCollam$^{1,2}$, G. A. de Wijs$^1$, and A. Fasolino$^1$}

\affiliation {
$^1$ Radboud University, Institute for Molecules and Materials, Heyendaalseweg 135, 6525 AJ Nijmegen, The Netherlands}

\affiliation {$^2$ High Field Magnet Laboratory (HFML-EMFL), Radboud University, 6535 ED Nijmegen, The Netherlands
         }

\begin{abstract}
Rashba spin splitting in two-dimensional (2D) semiconductor systems is generally calculated in a ${\bf k} \cdot {\bf p}$ Luttinger-Kohn  approach where the spin splitting due to asymmetry emerges naturally from the bulk band structure. In recent years, several new classes of 2D systems have been discovered where electronic correlations are believed to have an important role. In these correlated systems, the effects of asymmetry leading to Rashba splitting have typically been treated phenomenologically. We compare these two approaches for the case of 2D electron systems in SrTiO$_3$-based heterostructures, and find that the two models produce fundamentally different behavior in regions of the Brillouin zone that are particularly relevant for magnetotransport.  Our results demonstrate the importance of identifying the correct approach in the quantitative interpretation of experimental data, and are likely to be relevant to a range of 2D systems in correlated materials.

\end{abstract}
\pacs{73.20.-r, 71.15.-m,71.20.-b,75.47.-m}
\maketitle

\section{Introduction}

\begin{figure*}
\begin{center}
\includegraphics[angle=0,width=0.5\textwidth]{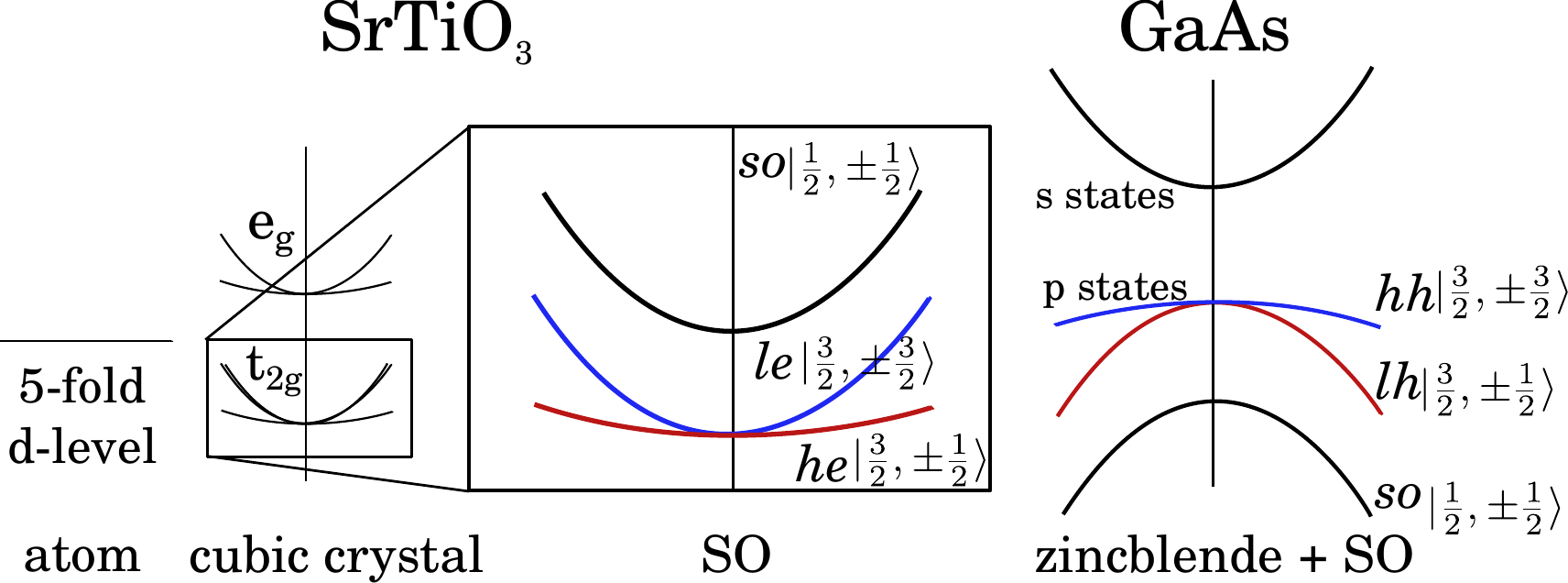}
\caption{Comparison of the conduction band $d$-states in SrTiO$_3$ with the hole band $p$-states in GaAs (see text). The color red (blue) indicates $m_j=\pm 3/2$ ($m_j=\pm 1/2$) bands to point out the reversal of the heavy and light mass.
}
\label{fig:2}
\end{center}
\end{figure*}

The discovery of a 2D electron system at the LaAlO$_3$/SrTiO$_3$ (LAO/STO) interface~\cite{ohtomo2004high} has generated a large interest in this topic, due to the realization of a new class of materials based on transition metal oxide heterostructures, and also to the diverse phenomena that they exhibit, such as superconductivity, magnetism, and strong spin-orbit interactions, all of which can be tuned by an external electric field~\cite{caviglia2008SC, ruhman2014, caviglia2010tunable,Stornaiuolo2014,Biscaras,Grilli}.  The development of heterostructures made from materials traditionally studied for correlated electron effects such as high-temperature superconductivity, Mott physics, and quantum magnetism,  invites us to inquire whether approaches that have been successful for semiconductor heterostructures can become relevant for this new class of materials~\cite{sulpizio2014nanoscale}.
A key connection to semiconductor physics comes through STO, 
where strong spin-orbit coupling deeply affects the free electron character of the bulk electronic bands, such that they can be theoretically described in ways analogous to the hole bands in III-V semiconductors~\cite{winkler} (see Fig.\ref{fig:2}).
Recently, many experimental results have addressed the Rashba spin splitting due to asymmetry at the interfaces in oxide heterostructures in presence of spin-orbit coupling~\cite{caviglia2010tunable,nakamura2012experimental,
fete2012rashba,fete2014large,ben2010tuning,liang2015nonmonotonically,
diez2014giant,king2014quasiparticle}. It is also important to consider the effect of unintentional asymmetry in nominally symmetric systems, such as $\delta$-doped materials~\cite{Bell}.
Here we compare the effect on a given conduction band structure of two descriptions of the Rashba spin-splitting that have been developed independently in the two fields of semiconductors and correlated electrons. Our results show a clear difference between the two descriptions that could be important for the interpretation of experimental results, in particular as a consequence of the different resulting Fermi surfaces. Our findings will also have relevance to other 2D systems in correlated materials where spin-orbit coupling and Rashba effects are important, such as topological insulators, iridium oxides and skyrmionic systems~\cite{Xu2014,Witczak,Biscaras,Lake,Banerjee}.

In the field of correlated electrons, it has become customary to assume that a suitable description of the  spin splitting due to asymmetry  can be obtained by adding to the Hamiltonian the following phenomenological term~\cite{petersen2000simple, zhong2013theory, khalsa2013theory,kim2013origin,diez2014giant,zhou2015spin}
\begin{equation}\label{eq:Hgamma}
 H^\gamma=\gamma\left(
 \begin{array}{ccc}
 0&0&2 i \text{sin}(k_x)\\
0&0&2 i \text{sin}(k_y)\\
-2 i \text{sin}(k_x) &-2 i \text{sin}(k_y) & 0
 \end{array}
 \right).
\end{equation} 
This model depends on the scalar parameter $\gamma$ which is a measure of the asymmetry of atomic orbitals at the interface.

Alternatively, within the ${\bf k} \cdot {\bf p}$ Luttinger Kohn (LK) method~\cite{luttinger1955motion}
 of spin orbit coupled bulk bands, the spin splitting follows naturally from the non-symmetric confining potential. 
This approach is well established for semiconductors~\cite{goldoni1992spin,andrada1994spin,shapers1998effect,
engels1997experimental,koga2002rashba,winkler} and has been shown to produce Rashba spin splittings in the LaAlO$_3$/SrTiO$_3$ heterostructure in Ref.~\onlinecite{vanHeeringen2013k.p}. 
Although both models, hereafter referred to as the $\gamma$-model and the LK-model, aim to capture the asymmetry of the system, the resulting spin splittings are in fact qualitatively and quantitatively different due to fundamental differences in the underlying Hamiltonians that we describe in the following. 
These differences can be crucial in the interpretation of magnetotransport measurements that measure the effect of Rashba spin splitting to extract details of the band structure~\cite{caviglia2010tunable,nakamura2012experimental,
fete2012rashba,fete2014large,ben2010tuning,liang2015nonmonotonically,
diez2014giant,king2014quasiparticle}.

It is still an open question which asymmetry is dominant in SrTiO$_3$-based heterostructures, a localized distortion of atomic orbitals at the interface or an effective asymmetric confining potential acting on more extended wavefunctions in the SrTiO$_3$~\cite{mannhart2010oxide,khalsa2013theory,vanHeeringen2013k.p}.
Measurements of the Rashba spin splitting can help to answer this question but the issue is complicated by the fact that access to  the Rashba spin splittings is usually indirect, {\it via} either weak anti-localization or Shubnikov de Haas (SdH) oscillations.

To make a correct identification attainable we compare the Rashba spin splittings as produced by the $\gamma$- and LK-model and identify the key differences.
We find that the two models  give comparable spin splitting of the lowest heavy electron states if the parameter $\gamma$ is suitably chosen.
At higher energies, in the vicinity of typical Fermi energies, the $\gamma$-model gives much larger spin splittings  than the LK model, with different wavevector dependence. These differences lead to different Fermi surfaces in the two models. This means that it would be possible to discriminate between them by SdH experiments~\cite{McCollam}  if the Fermi energy is swept, e.g. by gating, through the energy range of the subband anticrossing and beyond, where the models differ significantly.  

In Sec.~II we describe the orbital character of the relevant electrons and compare them to the hole states in III-V semiconductors where the LK model is routinely used for calculations of electronic structure. In Sec.~III we discuss the origin of the Rashba splitting  for SO coupled bands in asymmetric potentials. In Section IV we show how the bulk bands are changed by SO coupling within the LK formalism. In particular we discuss how the $d_{xy}$, $d_{yz}$ and $d_{xz}$  states transform in presence of SO coupling to the light electrons ($le$), heavy electrons ($he$)  and  spin-off ($so$) bands. We also introduce the LK coupling hamiltonian $H^{\rm LK}_{\pmb{\uparrow\downarrow}}$. In Section V, we introduce the $H^{\gamma}_{\pmb{\uparrow\downarrow}}$ for the $\gamma$-model and we compare it to  $H^{\rm LK}_{\pmb{\uparrow\downarrow}}$ in Section VI. Section VII is devoted to a comparison of the results of the two models for the Rashba spin splitting. Conclusions and perspectives are given in Section VIII.

\section{Character of SO-coupled bands}
\label{sec:character}
The conduction band edge of SrTiO$_3$ has strong similarities to the valence band of III-V semiconductors and formally the same Luttinger Kohn (LK) Hamiltonian (Eq.~\ref{eq:cubicH}) can be used to describe quantitatively the conduction electrons. In the III-V semiconductors, this Hamiltonian combined with the envelope function method has been very successful in the study of heterostructures.~\cite{winkler} 
That a material where the conduction band consists of $d$-level states turns out to be similar to that of $p$-level valence states in III-V semiconductors may seem surprising. 
In Fig.~\ref{fig:2} we sketch the relation between the conduction band in SrTiO$_3$ and the valence band in III-V semiconductors.
In SrTiO$_3$ the cubic crystal field splits the fivefold degenerate atomic $d$-levels (tenfold counting spin) into the threefold degenerate $t_{2g}$ states ($d_{xy}$, $d_{yz}$, and $d_{zx}$) and the twofold degenerate $e_g$ states.
At $\Gamma$, the symmetry of the $t_{2g}$ states is $\Gamma_{25}^+$, the same as for the \textit{p}-states in the zincblende structure of III-V semiconductors~\cite{dresselhaus2008group,vanHeeringen2012scriptie}.
SO coupling lifts the degeneracy further into a heavy and light electron band (\textit{he, le}) corresponding to a total angular momentum $J=3/2$ and a split off band (\textit{so}) with $J=1/2$, analogous to the heavy and light holes (\textit{hh, lh}) and split off holes (\textit{so}) in III-V semiconductors. A noticeable difference is that the nature of the basis set (\textit{d} or \textit{p}) makes that the orbital character of the \textit{light} electrons corresponds to the \textit{heavy} holes and vice versa as indicated by the colors in Fig.~\ref{fig:2}.
We emphasize that due to strong SO coupling in SrTiO$_3$ labeling electrons as $d_{xy}$ or $d_{yz}$ states is not correct, and classification in terms of the total angular momentum $J$ is necessary.

\section{Rashba spin splitting}

In bulk systems and symmetric heterostructures bands are spin degenerate (in absence of a magnetic field).
In fact, space inversion symmetry  $E(k,\uparrow)=E(-k,\uparrow)$ combined with time reversal symmetry (Kramers degeneracy) 
$E(k,\uparrow)=E(-k,\downarrow)$ yields  $E(-k,\uparrow)=E(-k,\downarrow)$.
If space reversal symmetry is broken, bands can spin split although they still obey Kramers degeneracy.
When such a spin splitting is a result of an asymmetric confining potential it is called Rashba spin splitting, named after Rashba who first described it for a simple 2-band model~\cite{bychkov1984properties}.
Today, the term Rashba spin splitting is also used to describe multiband systems such as SrTiO$_3$-based heterostructures~\cite{winkler}.

As discussed in Sec.~\ref{sec:character} the relevant electrons occupy the $J=3/2$ and $J=1/2$ multiplets for which we use the following notation:
\begingroup
\renewcommand*{\arraystretch}{1.5}
\begin{equation}\label{eq:Jbasis}
\begin{array}{c}
\pmb{\uparrow}= \left\{
\begin{array}{l}
\ket{3/2,+3/2}:= le \uparrow\\
\ket{3/2,-1/2}:= he\uparrow\\
\ket{1/2,-1/2}:= so\uparrow
\end{array}\right. \\
\pmb{\downarrow} = \left\{
\begin{array}{l}
\ket{3/2,-3/2}:= le\downarrow\\
\ket{3/2,+1/2}:= he\downarrow\\
\ket{1/2,+1/2}:= so\downarrow
\end{array}\right.
\end{array}
\end{equation}
\endgroup
We give these states in terms of $d_{xy}$, $d_{yz}$, and $d_{zx}$ as Supplemental Material S1. 
Since we focus on the interaction and splitting of the $\pmb{\uparrow}$ and $\pmb{\downarrow}$ states we write the Hamiltonian as block matrix: 
\begingroup
\renewcommand*{\arraystretch}{1.5}
\begin{equation}\label{eq:H}
H=\left(
\begin{array}{cc}
H_{\pmb{\uparrow}\pmb{\uparrow}} & H_{\pmb{\uparrow}\pmb{\downarrow}}\\
H_{\pmb{\downarrow}\pmb{\uparrow}} & H_{\pmb{\downarrow}\pmb{\downarrow}}
\end{array}
\right).
\end{equation}
\endgroup
The diagonal blocks  $H_{\pmb{\uparrow\uparrow}}$ and $H_{\pmb{\downarrow\downarrow}}$ give the spin degenerate bands. 
The interaction between the $\pmb{\uparrow}$ and the $\pmb{\downarrow}$ states is governed by the off-diagonal blocks $H_{\pmb{\uparrow\downarrow}}$ en $H_{\pmb{\downarrow\uparrow}}$.
In  Sec.~\ref{sec:LKmodel} and \ref{sec:gammamodel} we cast the LK Hamiltonian and the $\gamma$-model in a $\ket{J,m_j}$ basis to allow a direct comparison of the coupling term $H_{\pmb{\uparrow\downarrow}}$.

\section{Determination of $H_{\uparrow\downarrow}$ in the LK formalism}
\label{sec:LKmodel}
As discussed in Sec.~\ref{sec:character}, near $\Gamma$ the $t_{2g}$ bands of cubic SrTiO$_3$ neglecting SO are described by the ${\bf k} \cdot {\bf p}$ Hamiltonian~\cite{dresselhaus2008group,bistritzer2011electronic, vanHeeringen2013k.p,vanHeeringen2012scriptie}.
\begin{widetext}
\begin{equation}\label{eq:cubicH}
 H_{cubic}\left(
\begin{array}{c}
d_{yz} \\
d_{zx} \\
d_{xy}\\
\end{array} 
 \right)=\left(
 \begin{array}{ccc}
 Lk_x^2+M(k_y^2+k_z^2)&Nk_xk_y & Nk_xk_z\\
  Nk_xk_y & Lk_y^2+M(k_x^2+k_z^2)& Nk_yk_z \\
  Nk_xk_z & Nk_yk_z & Lk_z^2+M(k_x^2+k_y^2)
 \end{array}
 \right)
 \left(
\begin{array}{c}
d_{yz} \\
d_{zx} \\
d_{xy}\\
\end{array} 
 \right)
 .
\end{equation}
\end{widetext}
For convenience, throughout this paper we write the wave vector $k$ in dimensionless units, i.e. multiplied by the lattice vector $a=3.905$~\AA$^{-1}$.
These bands are all two-fold degenerate when spin is considered.
The effective mass parameters $L=40.03$~meV, $M=638$~meV and $N=105.9$~meV  and the SO splitting $\Delta_{\rm SO}=28.5$~meV have been determined for SrTiO$_3$in Ref.~\onlinecite{vanHeeringen2013k.p} by comparison to density functional theory (DFT) calculations. 

\begin{figure}
\begin{center}
\includegraphics[angle=0,width=0.5\textwidth]{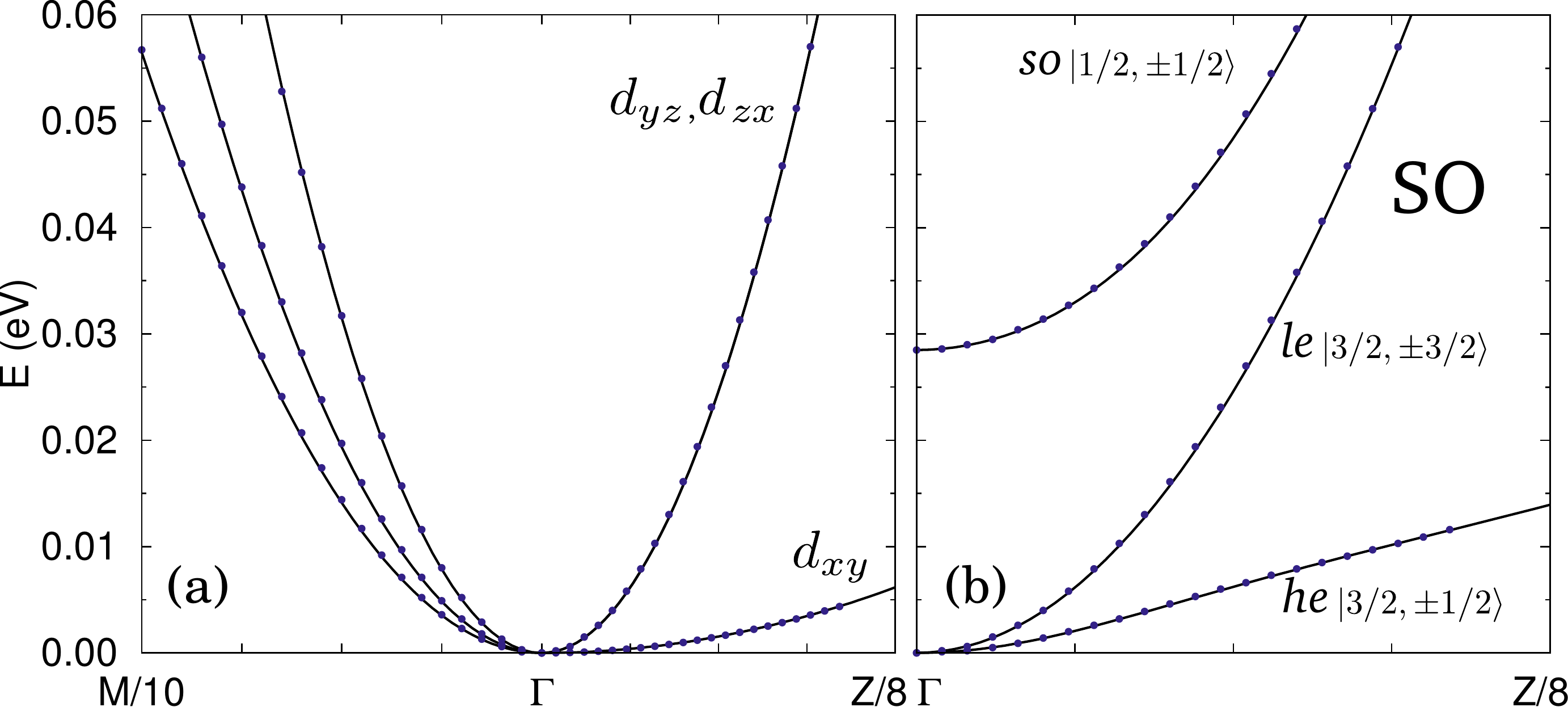}
\caption{${\bf k} \cdot {\bf p}$ model (lines) and DFT calculations~\cite{vanHeeringen2013k.p} (dots) of the band structure of bulk SrTiO$_3$ around $\Gamma$. Left panel: no SO, Eq.~\ref{eq:cubicH}. The non degenerate bands along $\Gamma-M$ require $N\neq0$ (see text). Right panel: with SO, Eq.~\ref{eq:Hlk}. 
}
\label{fig:3}
\end{center}
\end{figure}

Fig.~\ref{fig:3}a shows the $t_{2g}$ ($d_{xy}$, $d_{yz}$, and $d_{zx}$) bulk bands according to this model in comparison with the DFT calculations.

We note that along the $(0,0,k_z)$ direction, all off-diagonal terms of Eq.~\ref{eq:cubicH} vanish, leaving the parabolic dispersion given by the diagonal elements. The $d_{yz}$ and $d_{zx}$ are degenerate with light effective mass ($\propto 1/M$) while  the $d_{xy}$ has a heavy effective mass ($\propto 1/L)$.
Moreover, if $N$ would be zero, all off-diagonal term would vanish also along the in-plane directions, with opposite character of the effective masses. For instance, the $d_{xy}$ band which is heavy along the $\Gamma-Z$ direction $(0,0,k)$ would have a light mass ($\propto 1/M)$ in the $\Gamma-M$ direction $(k,k,0)$.
We note, however,  that, along $\Gamma-M$, the three bands are non degenerate~\cite{vanHeeringen2013k.p,janotti2011strain} which requires $N\ne0$, although recent experiments suggest $N=0$~\cite{allen2013conduction}. This fact is important for the evaluation of the Rashba splitting as we show below.  

The SO interaction is added to the Hamiltonian. In the $\ket{J,m_j}$ basis the additional term is diagonal:
\begin{equation}
H_{\rm SO}=\frac{\Delta_{\rm SO}}{3}\, \text{diag}[-1,-1,-1,-1,2,2].
\end{equation}
The total Hamiltonian is given by
\begin{equation}
\label{eq:Hlk}
H^{\rm LK}=H_{\rm cubic} + H_{\rm SO}.
\end{equation}
Since our main aim is to establish the appropriate model to describe the STO spin-orbit coupled bands, in this study  we neglect  the small tetragonal distortion occurring below $T=110$~K~\cite{mattheiss1972effect} and refer to Ref.~\onlinecite{vanHeeringen2013k.p} for more details. 

The bands calculated according to Eq.~\ref{eq:Hlk} with $\Delta_{\rm SO}=28.5$~meV  are shown in Fig.~\ref{fig:3}b in comparison to DFT calculations. In our calculations we shift the energy origin by $\Delta_{\rm SO}/3$ so that the band edge of the $J=3/2$ multiplet remains at $E=0$.  The SO interaction lifts the degeneracy of the $d_{yz}$  and $d_{zx}$ bands. One of them is lifted up in energy by $\Delta_{\rm SO}$ becoming the split-off $so|1/2,\pm1/2\rangle$ while the other becomes the pure $le|3/2,\pm3/2\rangle$ keeping the same light effective mass ($\propto 1/M$).  The $so$  band is coupled to the $he|3/2,\pm1/2\rangle$ band and the increased energy separation reduces  the $d_{xy}$ effective mass $m_{xy}=6.2~m_e$ to $m_{he}=1.2~m_e$. Most of the literature about the Rashba splitting still describes the results in terms of the $d$ states, which is, strictly speaking, not correct in presence of SO coupling. Nevertheless, qualitatively, the feature of two bands, one with very heavy mass and another with lighter mass in the quantization direction and opposite behavior in the in-plane direction, leading to subband crossing,
 is common to the two descriptions.  The effect of a quantizing potential is further described in Section~VII.

When written in the $\ket{J,m_j}$ basis the Hamiltonian is of the form of Eq.~\ref{eq:H}. The explicit form of $H_{\pmb{\uparrow\uparrow}}$ and $H_{\pmb{\downarrow\downarrow}}$ is given in Section S2 of the Supplemental Material. Here we focus on the off-diagonal blocks that  couple the $\pmb{\uparrow}$ and $\pmb{\downarrow}$ states:
\begingroup
\renewcommand*{\arraystretch}{1.5}
\begin{equation}
\label{eq:HLKupdown}
 H_{\pmb{\uparrow\downarrow}}^{LK} 
\left(
\begin{array}{c}
le \downarrow \\
he \downarrow \\
so \downarrow\\
\end{array}
\right) 
 = Nk_z\left(
\begin{array}{ccc}
0  & \frac{-i}{\sqrt3} k_- & \frac{1}{\sqrt 6} k_-\\
\frac{i}{\sqrt3} k_- & 0 & \frac{1}{\sqrt2} k_+ \\
\frac{1}{\sqrt 6} k_- & \frac{1}{\sqrt2} k_+ & 0
\end{array}
\right)
\left(
\begin{array}{c}
le \downarrow \\
he \downarrow \\
so \downarrow\\
\end{array}
\right) 
\end{equation}
\endgroup
with $k_\pm =k_x \pm ik_y$. 

This matrix is nonzero also in bulk SrTiO$_3$ but does not lead to any spin splitting there, as it should not by spatial and time inversion symmetry. 
In heterostructures however, $k_z$ is an operator and $H_{\pmb{\uparrow\downarrow}}$ splits $\pmb{\uparrow}$ and $\pmb{\downarrow}$ states in absence of spatial inversion symmetry, resulting in a Rashba spin splitting proportional to $N$.

\section{$\gamma$-model in the $\ket{J,m_j}$ basis}
\label{sec:gammamodel}
In Refs.~\onlinecite{khalsa2013theory} and \onlinecite{zhong2013theory} where $N$ is assumed to be zero, the phenomenological term  $H^{\gamma}$ (Eq.~\ref{eq:Hgamma}) is added to a tight binding Hamiltonian to account for the extra hopping terms between the $d_{xy}$ and $d_{yz},d_{zx}$ states due to deformation of the atomic orbitals by the internal electric field at the interface~\cite{petersen2000simple,zhong2013theory,khalsa2013theory}.
We study the effect of this type of Rashba spin splitting within the envelope function method defining the $\gamma$ model as $H^{\rm LK}(N=0)+H^\gamma$.

Here we write $H^{\gamma}$ (Eq.~\ref{eq:Hgamma}) in the $\ket{J,m_j}$ basis to allow a direct comparison with $H_{\pmb{\uparrow\downarrow}}^{\rm LK}$. Using the approximation $\sin(k_x)\approx k_x$ we find
\begingroup
\renewcommand*{\arraystretch}{1.5}
\begin{equation}
\label{eq:Hgammaupdown}
H_{\pmb{\uparrow\downarrow}}^\gamma
\left(
\begin{array}{c}
le \downarrow \\
he \downarrow \\
so \downarrow\\
\end{array}
\right) 
= 2 i \gamma \left(
\begin{array}{ccc}
0  & -\frac{i}{\sqrt3} k_- & \frac{1}{\sqrt6} k_-\\
-\frac{i}{\sqrt3} k_- & -\frac{2i}{3} k_+ & -\frac{1}{3\sqrt2} k_+ \\
-\frac{1}{\sqrt6} k_- & \frac{1}{3 \sqrt2 } k_+ & \frac{2i}{3}k_+
\end{array}
\right)
\left(
\begin{array}{c}
le \downarrow \\
he \downarrow \\
so \downarrow\\
\end{array}
\right) 
\end{equation}
\endgroup

\section{Comparisons of $H_{\pmb{\uparrow\downarrow}}$ in the $\gamma$- and LK-model}
\label{sec:comparison}
In this section we compare the effect of the coupling matrix $H_{{\uparrow\downarrow}}$ in the $\gamma$- and LK-models on the spin splitting in an asymmetric confining potential. Despite the similar structure, the two coupling matrices Eq.~\ref{eq:HLKupdown} and \ref{eq:Hgammaupdown} present important differences. First, $H_{\pmb{\uparrow\downarrow}}^{\rm LK}$ is proportional to $k_z$ which in a confining potential becomes an operator whereas $H_{\pmb{\uparrow\downarrow}}^\gamma$ is proportional to the scalar $\gamma$. Second, each imaginary component in the elements of one matrix is real in the other and vice versa. Third and most important, not all diagonal elements of $H_{\pmb{\uparrow\downarrow}}^{\gamma}$ are zero, giving a direct coupling between \textit{he}$\uparrow$ and \textit{he}$\downarrow$ and between \textit{so}$\uparrow$ and \textit{so}$\downarrow$. It is this feature that eventually leads to the much larger spin splitting of the lowest subband that we describe in the next section \ref{Sec:Numerical}.

Close to $\Gamma$, the $\gamma$-model gives linear $k$ splitting for {\it he} subbands, and $k^3$ splitting for {\it le} subbands.
For the LK-model, due to the zero valued diagonal elements, it is the mixed {\it so, he} character of the lowest bulk band that gives rise to a linear splitting. In fact, the $he$ and $so$ states are coupled by terms in $k_zk_\pm$ and $k_z^2$ that couple states of different space parity. When the parity of the states is broken by the electric field, the coupling occurs also with the same subband and results in a linear spin orbit splitting. The pure {\it le} bulk band, instead, has a $k^3$ splitting due to a two-step coupling.
 
Due to the different coupling mechanism in the two models, a single value of $\gamma$ can reproduce only either the linear or the cubic splitting for a given electric field.
We found that fitting $\gamma$ to the linear splitting gives a much better overall agreement (whereas fitting of the cubic splitting largely overestimates the splittings). Therefore we identify for every value of the electric field, a parameter $\gamma_{\rm fit}$ that gives the same linear spin splitting of the lowest subband in both models. As shown in detail in Section S4 of Supplemental Material, $\gamma_{\rm fit}$ grows with electric field. This finding is in agreement with the observation of increased SO coupling with carrier concentration discussed in Ref.\onlinecite{Stornaiuolo2014}.

\section{Numerical evaluation: comparison of the Rashba spin splitting in LK-model and the $\gamma$-model}
\label{Sec:Numerical}
\begin{figure}[ht!]
\includegraphics[angle=0,width=0.5\textwidth]{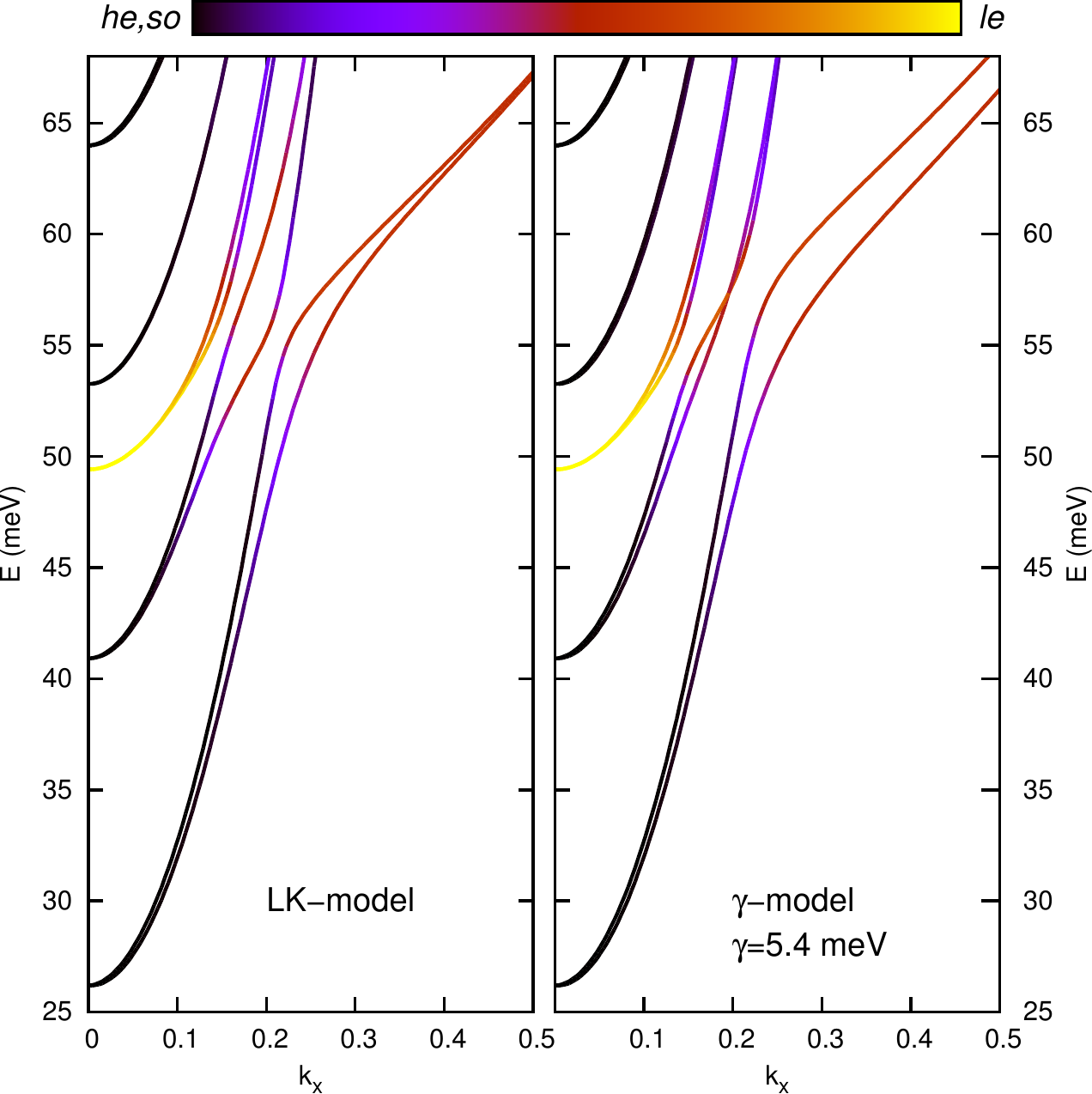}
\caption{In-plane dispersion in a triangular well with slope $F=1.0$~meV/\AA\ according to the LK Hamiltonian (left panel) and the $\gamma$-model with $\gamma=5.4$~meV (right panel). The color indicates the {\it he, so} or {\it le} character of the subbands.
}
\label{fig:4}
\end{figure}
Here we compare the Rashba spin splitting of the LK- and the $\gamma$-model. 

We consider a triangular confining potential $V(z)=Fz$ resulting from a constant electric field with electric field strength $F$. We find the quantized subbands by solving the eigenvalue problem by finite differences as described in Ref.~\onlinecite{vanHeeringen2013k.p} and in Section S3 of the Supplemental Material. 
Self consistent calculations of the potential would lead  to a bending at high energies with negligible effects on the lowest subbands, except in the situation where the Fermi energy comes close to the upper edge of the confining potential~\cite{Bergeal}, where it could lead to electronic instabilities~\cite{Scopigno}. For simplicity, we neglect these effects here. 

A triangular potential has the advantage that $\gamma$ depends only on the electric field strength and not on the spatial coordinate $z$.
Our method is based on a ${\bf k} \cdot {\bf p}$ description of the bulk bands around $\Gamma$ which becomes less accurate for large confinement energies limiting the electric field strength to $F\sim 1.0$ meV/\AA.

We point out that the envelope function method gives all the quantized subbands. For example, it is possible to have more than one subband with the same character at $\Gamma$, as can be seen in Fig.~\ref{fig:4}. In other studies instead~\cite{zhong2013theory,diez2014giant} only one {\it he}, one {\it le} and one {\it so} subband are calculated and the quantization energy is chosen as a parameter $\Delta_I$. 

In Fig.~\ref{fig:4} we show the subbands calculated within the two models for  $F=1.0$ meV/\AA~where two {\it he} subbands are lower than the first {\it le} subband, as indicated by the color coding.
One can see that the {\it le} subband, which has a heavy mass in the $xy$- plane, anticrosses the two {\it he} subbands at $E\sim55$~meV. At this anticrossing, which has also been identified as a  Lifshitz point~\cite{joshua2012universal,diez2014giant,ruhman2014competition}, the Rashba spin splitting is largest in both models.
For this electric field strength, at $\Gamma$, we find that the lowest $he$ subband is $\approx 22$ meV below the lowest $le$ subband, to be compared to $\sim60-90$~meV~\cite{santander2011two,joshua2012universal,salluzo2009orbital,cancellieri2014doping} and much larger splittings of $>200$~ meV~\cite{king2014quasiparticle,plumb2014mixed,zhong2013theory,
delugas2011spontaneous} found in other work.

Up to the anticrossing both models give very similar subbands, as guaranteed by the choice of $\gamma_{\text{fit}}$. At larger energies, the subbands start to mix producing clear differences between the two models. The most important difference is that the Rashba spin splitting of the lowest subband after the anticrossing decreases in the LK model, whereas it remains large in the $\gamma$-model. The splitting of the second band is also quite different in the two models, with a crossing in the $\gamma$-model. 
Since the experimental evidence of Rashba spin splitting comes mainly from transport measurements~\cite{liang2015nonmonotonically,fete2014large,
caviglia2010tunable,fete2012rashba,liang2015nonmonotonically,
nakamura2012experimental,ben2010tuning}, such a different dispersion at large $k$ is highly significant for the interpretation of experimental data. SdH experiments should be able to distinguish between the resulting different shapes of the Fermi surface, if the Fermi energy could be changed continuously by means of external gating. Moreover, the different composition of the subband eigenvectors affects the spin orientation and would lead to different spin textures, as described in section 6.6 of Ref.\onlinecite{winkler}.     

\begin{figure*}[ht!]
\begin{center}
\includegraphics[angle=0,width=1.0\textwidth]{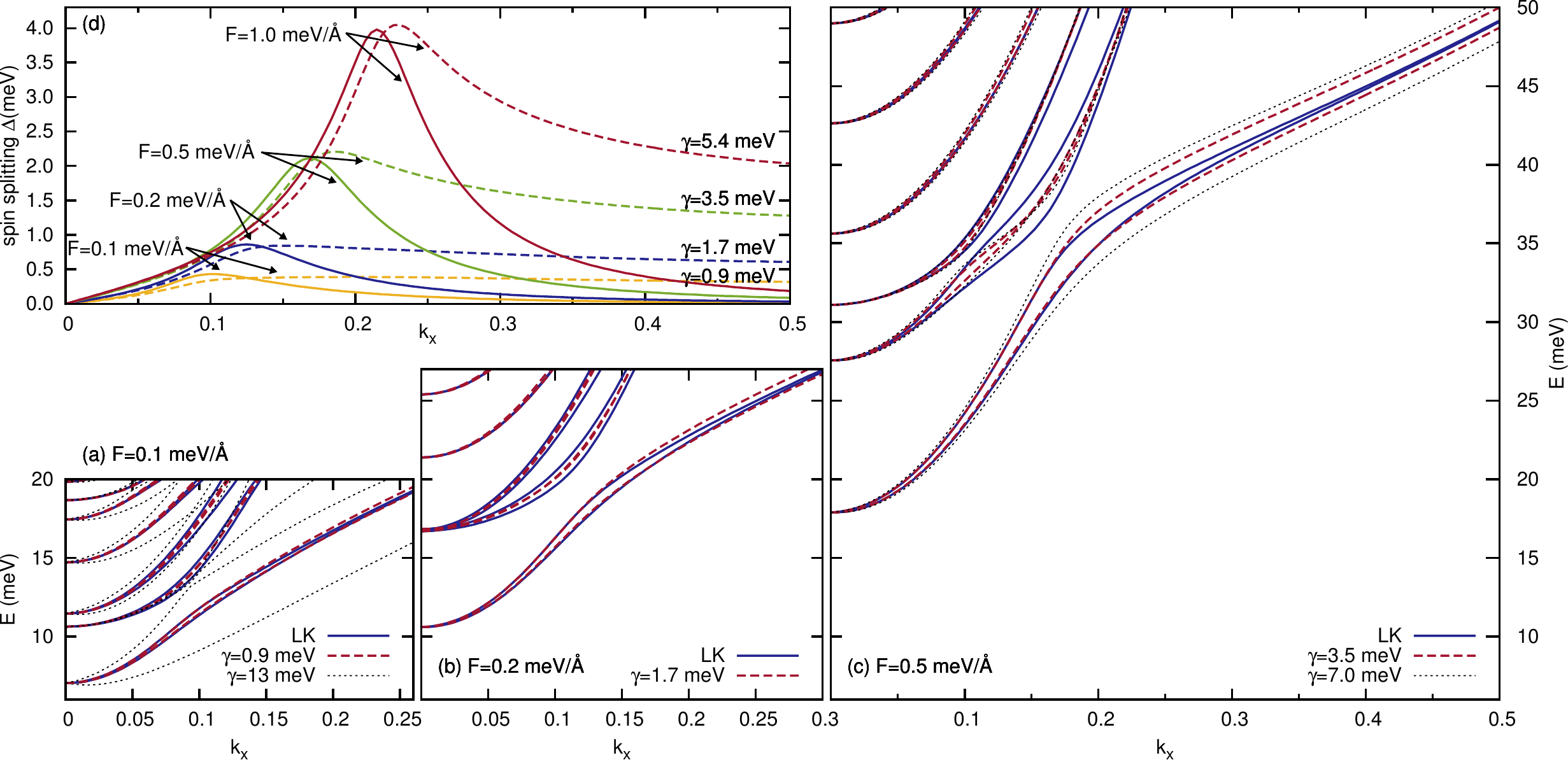}
\caption{Subbands for $F=0.1$~meV/\AA~(a), $0.2$~meV/\AA~(b), $0.5$~meV/\AA~(c) and spin splitting $\Delta$ of the lowest subband for four values of $F$ (d). Solid line is the LK-model, and the dashed line the $\gamma$-model with $\gamma_{\text{fit}}$. The dotted lines in (a) and (c) correspond to larger values of $\gamma$. In Fig.~\ref{fig:55}d we show the spin splitting of the lowest subband.
The spin splitting is very sensitive to the choice of the parameter $\gamma$. 
In panel Fig.~\ref{fig:55}a we show the subbands for $\gamma=13$~meV which is the value that gives the same cubic splitting of the {\it le} (see Supplemental Material S4) and in Fig.~\ref{fig:55}c we show the effect of doubling $\gamma$.
}
\label{fig:55}
\end{center}
\end{figure*}
In Fig.~\ref{fig:55} we compare the two models for three different values of the electric field $F$. As $F$ increases, the potential well becomes steeper, and the subbands move upwards in energy to an extent that depends on their respective effective masses.  Due to the non parabolicity of the SO coupled bulk bands~\cite{vanHeeringen2013k.p}, the order of the bands changes from {\it he$_1$, le$_1$, he$_2$} for $F=0.1$ and $0.2$~meV/\AA~ to {\it he$_1$, he$_2$, le$_1$} for $F=0.5$~meV/\AA\ as for $F=1.0$~meV/\AA~ shown in Fig.~\ref{fig:4}. 
Other than this difference the main features remain unchanged: the two models agree well for energies well under the anticrossing, but the splitting is fundamentally different at higher energies. The LK model gives a splitting at every anticrossing.  The size of the splitting grows with the electric field because the anticrossing occurs at larger and larger values of the momentum, which determines the amount  of coupling. Moreover the splitting decreases away from the anticrossing. The $\gamma$ model, on the other hand, gives splitting only at the lowest anticrossing and the splitting remains large away from the anticrossing, due to the direct coupling of $he_\uparrow$ and $he_\downarrow$ described in Section VI. We emphasize again that the spin splitting is in the range of resolution of SdH experiments~\cite{McCollam}. For instance, for $F=0.5$ meV/\AA~ and a Fermi energy of 35 meV, one should be able to resolve the spin splitting of at least two of the three subbands crossing the Fermi energy.

\section{Summary and perspectives}
\label{sec:S&C}\

In summary, we have examined the Rashba spin splitting in STO-based heterostructures. We have compared in detail the results of the Luttinger-Kohn effective mass model of the $t_{2g}$ edge with a phenomenological model of the spin splitting inspired by the two band model of Rashba. The latter has the advantage of being simple to implement but has a weaker physical basis than the ${\bf k}\cdot {\bf p}$ model, for which spin splitting in asymmetric confining potentials results naturally, without addition of phenomenological parameters. We have pointed out that a discussion of the Rashba spin splitting in asymmetric potentials require to go over from a description in terms of the $d$ orbitals to the SO coupled $t_{2g}$ states, a basis which is intrinsic to the LK model. We have shown that the LK model predicts that spin splitting arises at the avoided crossing of the SO coupled subbands, 
and decreases away from them.  The $\gamma$ model has a very different, possibly unphysical, behaviour in this range. Establishing the right model to describe the Rashba spin splitting in oxide materials is important for a whole class of new 2D systems~\cite{Xu2014,Witczak,Banerjee} in correlated materials. 

\section*{Acknowledgements}
This work is part of the research program of the Stichting voor Fundamenteel Onderzoek der Materie (FOM), which is financially supported by the Nederlandse Organisatie voor Wetenschappelijk Onderzoek (NWO). AF and AM thank M. Salluzzo and AF thanks C. Noguera for fruitful discussions.

\end{document}



\title{Theoretical models of Rashba spin splitting in asymmetric  SrTiO$_3$-based heterostructures}

\author{L. W. van Heeringen, A. McCollam, G. A. de Wijs, and A. Fasolino}

\affiliation {
Radboud University, Institute for Molecules and Materials and High Field Magnet Laboratory, Heyendaalseweg 135, 6525 AJ Nijmegen, The Netherlands
         }
\setcounter{table}{0}
\setcounter{section}{0}
\setcounter{figure}{0}
\setcounter{equation}{0}
\renewcommand{\thepage}{\Roman{page}}
\renewcommand{\thesection}{S\arabic{section}}
\renewcommand{\thetable}{S\arabic{table}}
\renewcommand{\thefigure}{S\arabic{figure}}
\renewcommand{\theequation}{S\arabic{equation}}

\renewcommand*{\citenumfont}[1]{S#1}
\renewcommand*{\bibnumfmt}[1]{[S#1]} 
\begin{abstract}
\end{abstract}
\maketitle

\section{$\ket{J,m_j}$ basis}
The $\ket{J,m_j}$ states are linear combinations of the $d_{xy},d_{yz}$ and $d_{zx}$ states. We denote ($d_{xy}$,$d_{yz}$,$d_{zx}$) as ($X,Y,Z$). The $\ket{J,m_j}$ states are defined as
\begin{equation}\label{eq:ubasis}
\begin{aligned}
\ket{3/2,+3/2} =le\uparrow &=\frac{1}{\sqrt2}\ket{X + i Y\uparrow} \\
\ket{3/2,-1/2}=he\uparrow &=\frac{1}{\sqrt{6}}\ket{X -iY\uparrow} + \sqrt{\frac{2}{3}}\ket{Z\downarrow}\\ 
\ket{1/2,-1/2}=so\uparrow &=-\frac{i}{\sqrt3}\ket{X-iY\uparrow}+\frac{i}{\sqrt3}\ket{Z\downarrow}\\
\ket{3/2,-3/2}=le\downarrow &=\frac{i}{\sqrt2}\ket{X - i Y\downarrow} \\ 
\ket{3/2,+1/2} =he\downarrow &=\frac{i}{\sqrt6}\ket{X +iY\downarrow} - i \sqrt{\frac{2}{3}}\ket{Z\uparrow}\\ 
\ket{1/2,+1/2} =so\downarrow &=\frac{1}
{\sqrt3}\ket{X+iY\downarrow}+\frac{1}{\sqrt3}\ket{Z\uparrow}. \\ 
\end{aligned}
\end{equation}

\section{LK Hamiltonian in $\ket{J,m_j}$ basis}
In the $\ket{J,m_j}$ basis the LK Hamiltonian including SO coupling is given by 
\begingroup
\renewcommand*{\arraystretch}{1.5}
\begin{equation}
H^{LK}=\left(
\begin{array}{cc}
H^{LK}_{\pmb{\uparrow}\pmb{\uparrow}} & H^{LK}_{\pmb{\uparrow}\pmb{\downarrow}}\\
H^{LK}_{\pmb{\downarrow}\pmb{\uparrow}} & H^{LK}_{\pmb{\downarrow}\pmb{\downarrow}}
\end{array}
\right)
\end{equation}
\endgroup

with

\begin{equation}
H^{LK}_{\pmb{\uparrow}\pmb{\uparrow}} = \left(
\begin{array}{ccc}
p                     & b & -i\sqrt2 b        \\
b^\dagger              & q & c^\dagger        \\
i\sqrt2 b^\dagger      & c & r
\end{array}\right)
\end{equation}

and

\begin{equation}
H^{LK}_{\pmb{\downarrow}\pmb{\downarrow}} = \left(
\begin{array}{ccc}
p         & b^\dagger & -i\sqrt2 b^\dagger    \\
b         & q & c^\dagger        \\
i\sqrt2 b      & c & r
\end{array}\right)
\end{equation}
where
\begin{equation}
\begin{aligned}
p=&\frac{1}{2}(L+M)(k_x^2+k_y^2)+Mk_z^2  \\
q=&\frac{1}{6}(L+5M)(k_x^2+k_y^2)+\frac{1}{3}(2L+M)k_z^2\\
r=&\frac{1}{3}(L+2M)(k_x^2+k_y^2+k_z^2)+\Delta_{SO}\\
b=&\frac{1}{2\sqrt{3}}(L-M)(k_x^2-k_y^2)-\frac{1}{\sqrt{3}}iNk_xk_y \\
c=&\frac{1}{3\sqrt2}i(L-M)(k_x^2+k_y^2-2k_z^2).\\
\end{aligned}
\end{equation}
$H^{LK}_{\pmb{\downarrow}\pmb{\uparrow}}$ can be constructed from $H^{LK}_{\pmb{\uparrow}\pmb{\downarrow}}$ given in 
Eq.~7 of the main text using the Hermiticity of $H^{LK}$ implying 
\begin{equation}
H^{LK,ij}_{\pmb{\downarrow}\pmb{\uparrow}}=
(H^{LK,ji}_{\pmb{\uparrow}\pmb{\downarrow}})^\dagger.
\end{equation}
The same holds for $H_{\pmb{\downarrow}\pmb{\uparrow}}^{\gamma}$ using 
Eq.~8 of the main text.
\section{Envelope function method}
In a heterostructure, a superimposed potential landscape leads to quantization of $k_z$. According to the envelope function method, the wave function can be written as 
\begin{equation}
 \psi_{k_{||},k_z}(\mathbf{r})= e^{i\mathbf{k}_{||}\cdot\mathbf{r}_{||}} \bm{u(\bm{r})}\cdot \bm{\phi}(z,k_{||})
\end{equation}
with  $\mathbf{k}_{||}=(k_x,k_y,0)$, $\mathbf{r}_{||}=(x,y,0)$ and the basis functions $u_i(\bm{r})$ have the periodicity of the bulk unit cell, and are the $\ket{J, m_j}$ states in this study.
To calculate the electronic dispersion in SrTiO$_3$-based heterostructures we calculate the components of $\bm{\phi}(z,k_{||})$ by adding the appropriate potential profile $V(z)$ to the Hamiltonian of 
Eq.~3, treating $k_z$ as $-i\hbar\frac{d}{dz}$ and solving the resulting eigenvalue problem:
\begin{equation}\label{eq:appHphiEphi}
 \left\{ H\left(k_{||}, k_z \rightarrow -i\frac{\partial}{\partial z}\right)+V(z) \mathbf{I} \right\} \bm{\phi}(z,k_{||})=E(k_{||})\bm{\phi}(z,k_{||})
\end{equation}
This equation is solved by the finite difference method by discretizing the envelope wave function on an equispaced grid in real space. 

\section{Choice of $\gamma$}

\begin{figure}[ht!]
\includegraphics[angle=0,width=0.5\textwidth]{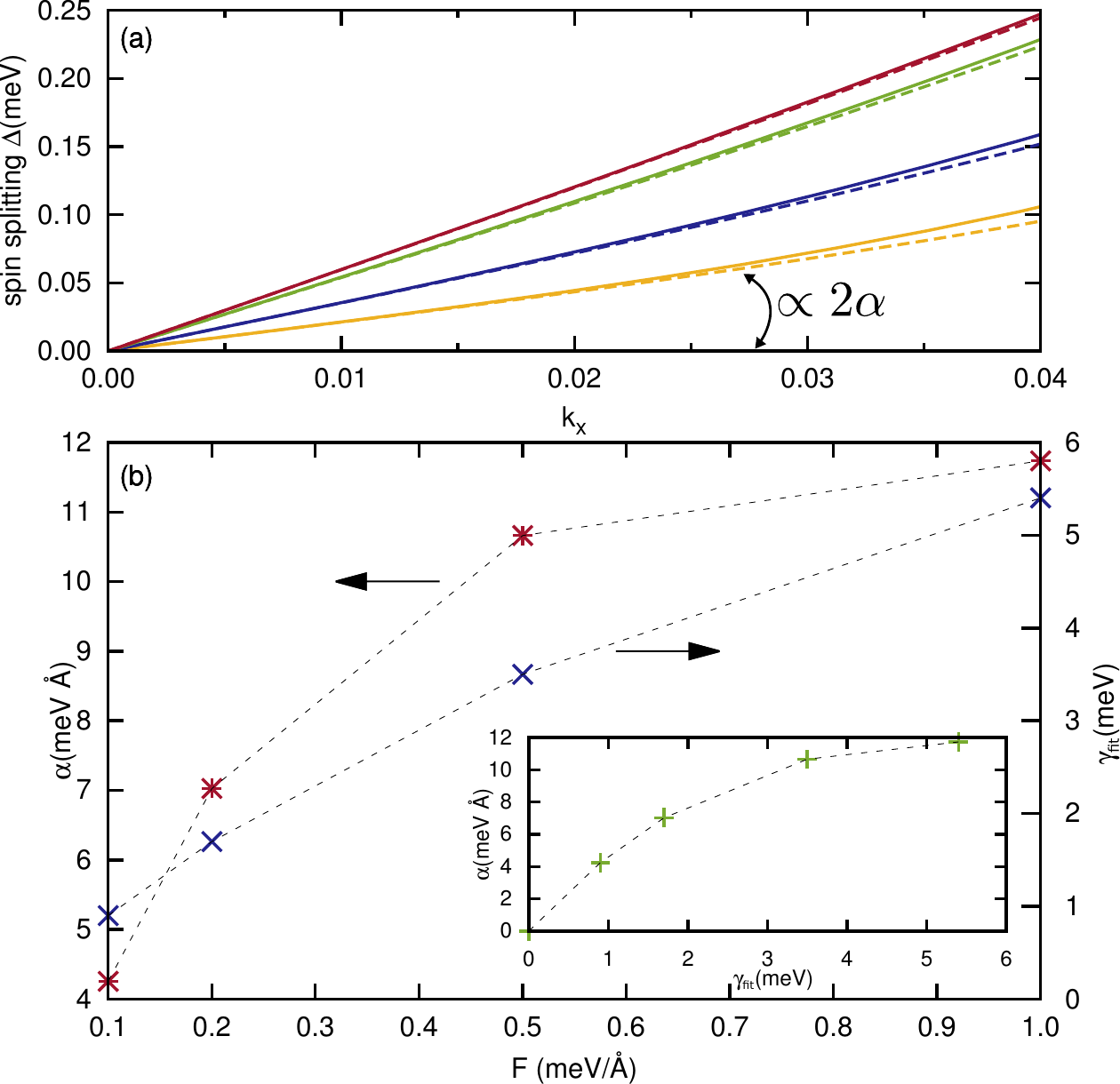}
\caption{(Color online) 
(a) Spin splitting of the lowest subband at small $k$ for $F=0.1$~meV/\AA~ (yellow), $F=0.2$~meV/\AA~ (blue), $F=0.5$~meV/\AA~ (green) and $F=1.0$~meV/\AA~ (red). Solid line is the LK-model, and the dashed line the $\gamma$-model with $\gamma_{\text{fit}}$. (b) $\gamma_{\text{fit}}$ (blue crosses) and $\alpha$ (red stars) as a function of $F$. Inset: $\alpha$ as a function of $\gamma_{\text{fit}}$ (green plus signs). $\gamma_{\text{fit}}$ is tuned to yield the same spin splitting of the lowest subband as the LK-model. $\alpha$ is determined from the slopes in (a). The dotted lines are guides to the eye.
}
\label{fig:6}
\end{figure}

The $\gamma$-model is based on the phenomenological, in principle unknown, parameter $\gamma$. To choose $\gamma$, for each value of the electric field $F$, we compare the LK and $\gamma$ model and choose $\gamma_{\text{fit}}$ as to give the same spin splitting of the lowest {\it he} subband around $\Gamma$.  In Fig.~\ref{fig:6}a we show, for several values of  $F$, the linear spin splitting $\Delta$ of the first subband close to $\Gamma$, calculated with the LK model (solid) and with the $\gamma$-model with $\gamma=\gamma_{\text{fit}}$. In the linear regime shown in Fig.~\ref{fig:6}a, one can define the Rashba coefficient $\alpha$ as  $\Delta=2 \alpha k$. The coefficient $\alpha$, which is a measure of the spin splitting, depends on the electric field strength
as  shown in Fig.~\ref{fig:6}b. The values of $\alpha$ are in the range found in other theoretical studies~\cite{zhong2013theory,khalsa2013theory} and experiments~\cite{king2014quasiparticle,caviglia2010tunable}, although also larger values  have been reported~\cite{fete2014large}. In
Fig.~\ref{fig:6}b we also show $\gamma_{\text{fit}}$ as a function of $F$. In fact, also $\gamma$ is a measure of the asymmetry and therefore expected to be proportional to $F$~\cite{petersen2000simple}. The growth of $\gamma$ with $F$ is in agreement with the observation of increased SO coupling with carrier concentration discussed in Ref.\onlinecite{Stornaiuolo2014}
In Fig.~3 and Fig.~4 of the main text, one can see that the choice of $\gamma$ obtained by fitting the splitting of the lowest {\it he} subband leads to a very good overall agreement of the two models below the anticrossing of {\it he} and {\it le} subbands.

\begin{figure}[ht!]
\includegraphics[angle=0,width=0.5\textwidth]{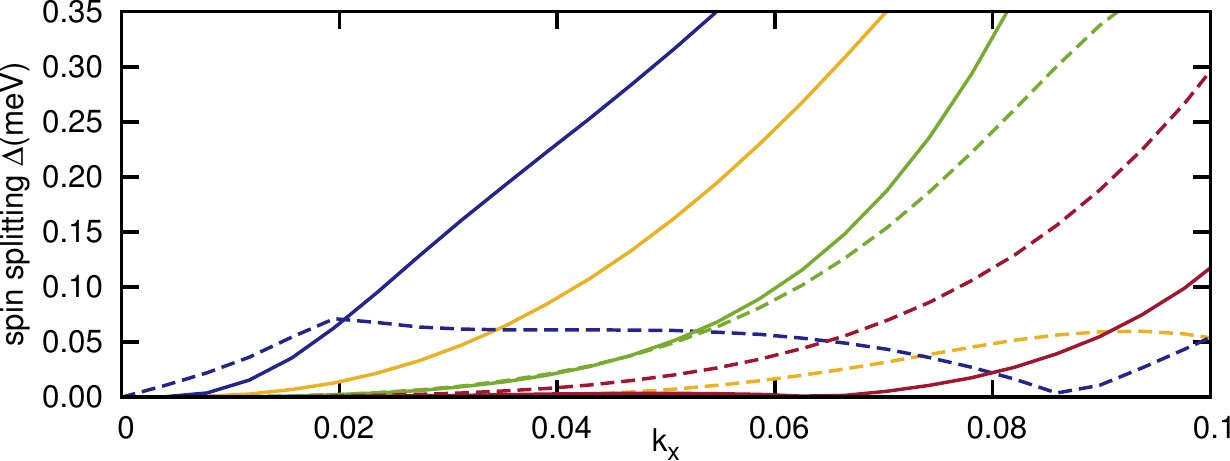}
\caption{(Color online) Spin-splitting of first {\it le} band for $F=0.1$~meV/\AA~ (yellow), $F=0.2$~meV/\AA~ (blue), $F=0.5$~meV/\AA~ (green) and $F=1.0$~meV/\AA~ (red). Solid line is the LK-model, and the dashed line the $\gamma$-model with $\gamma_{\text{fit}}$. 
}
\label{fig:7}
\end{figure}

In Fig.~\ref{fig:7} we compare also the spin splitting of the lowest {\it le} subband  given by the L-K model and by the $\gamma$
 model with $\gamma_{\text{fit}}$. This splitting is much smaller than that of the {\it he} subbands and  has a cubic behavior as
 anticipated in 
section IV of the main text and in agreement with other theoretical studies~\cite{zhong2013theory,kim2013origin} and
 found experimentally~\cite{nakamura2012experimental,liang2015nonmonotonically}. 
Only for $F=0.5$ meV/\AA~ $\gamma_{\text{fit}}$ gives a good agreement with the LK-model.

